\documentstyle[12pt,epsfig,amsbsy,amssymb,cite]{article}
 
\begin{document}
\begin{titlepage}

\begin{center}
  {\large   EUROPEAN ORGANIZATION FOR NUCLEAR RESEARCH}
\end{center}
\bigskip

\begin{flushright}
       CERN-EP-2003-082   \\ 
       28 November 2003
\end{flushright}
\bigskip\bigskip\bigskip\bigskip\bigskip

\begin{center}
  {\huge\bf 
Measurement of the partial widths of the Z into up- and down-type quarks
}
\end{center}
\bigskip\bigskip

\begin{center}
{\LARGE The OPAL Collaboration}
\end{center}

%
%
\begin{abstract}

Using the entire OPAL LEP1 on-peak Z hadronic decay sample,
$\rm Z\rightarrow q\bar{q}\gamma$ decays were selected by tagging 
hadronic final states with isolated photon candidates
in the electromagnetic calorimeter. 
Combining the measured 
rates of $\rm Z\rightarrow q\bar{q}\gamma$ decays with the total rate 
of hadronic Z decays permits the simultaneous determination of the 
widths of the Z into up- and down-type quarks. The values obtained, with total errors, were 
\begin{eqnarray}
  \Gamma_{\rm u}= 300 ^{+19}_{-18}\quad{\rm MeV}
  \qquad {\rm and} \qquad 
  \Gamma_{\rm d}= 381 ^{+12}_{-12}\quad{\rm MeV}.\nonumber
\end{eqnarray}
The results are in good agreement with the 
Standard Model expectation.

\end{abstract}
\vspace{15mm}
{\center{\large To be submitted to Physics Letters \\}}
\vspace{5mm}
%
\end{titlepage}

\newpage
\begin{center}{\Large        The OPAL Collaboration
}\end{center}\bigskip
\begin{center}{
G.\thinspace Abbiendi$^{  2}$,
C.\thinspace Ainsley$^{  5}$,
P.F.\thinspace {\AA}kesson$^{  3,  y}$,
G.\thinspace Alexander$^{ 22}$,
J.\thinspace Allison$^{ 16}$,
P.\thinspace Amaral$^{  9}$, 
G.\thinspace Anagnostou$^{  1}$,
K.J.\thinspace Anderson$^{  9}$,
S.\thinspace Arcelli$^{  2}$,
S.\thinspace Asai$^{ 23}$
% ...
}\end{center}\bigskip
\bigskip
$^{  1}$School of Physics and Astronomy, University of Birmingham,
Birmingham B15 2TT, UK
\newline
$^{  2}$Dipartimento di Fisica dell' Universit\`a di Bologna and INFN,
I-40126 Bologna, Italy
\newline
$^{  3}$Physikalisches Institut, Universit\"at Bonn,
D-53115 Bonn, Germany
\newline
% ...
$^{  5}$Cavendish Laboratory, Cambridge CB3 0HE, UK
\newline
% ...
$^{  9}$Enrico Fermi Institute and Department of Physics,
University of Chicago, Chicago IL 60637, USA
\newline
% ...
$^{ 16}$Department of Physics, Schuster Laboratory, The University,
Manchester M13 9PL, UK
\newline
% ...
$^{ 22}$Department of Physics and Astronomy, Tel Aviv University,
Tel Aviv 69978, Israel
\newline
$^{ 23}$International Centre for Elementary Particle Physics and
Department of Physics, University of Tokyo, Tokyo 113-0033, and
Kobe University, Kobe 657-8501, Japan
\newline
% ...
$^{  y}$ now at CERN
\newline
% ...
 
\newcommand{\epem}{\mbox{$\mathrm{e^+e^-}$}}
\newcommand{\Zzero}{\mbox{${\mathrm{Z}^0}$}}
\newcommand{\etal}{\mbox{{\it et al.}}}
\newcommand{\Opal}{\mbox{\rm OPAL}}
\newcommand{\LEP}{\mbox{LEP}}

\newcommand{\ee}{$\mbox{e}^+\mbox{e}^-$}
\newcommand{\e}{$\eta$}
\newcommand{\x}{$\xi$}
\newcommand{\z}{$\zeta$}
\newcommand{\ct}{$\cos\theta^{\bullet}$}
\newcommand{\mct}{\cos\theta^{\bullet}}
\newcommand{\mcct}{\cos^2\theta^{\bullet}}
\newcommand{\tb}{$\theta^{\bullet}$}
\newcommand{\bq}{\begin{equation}}
\newcommand{\eq}{\end{equation}}
\newcommand{\ba}{\begin{eqnarray}}
\newcommand{\ea}{\end{eqnarray}}
\newcommand{\bi}{\begin{itemize}}
\newcommand{\ei}{\end{itemize}}
\newcommand{\bn}{\begin{enumerate}}
\newcommand{\en}{\end{enumerate}}

\newcommand{\ycut}{\ifmmode{y_{\rm cut}} 
                   \else{$y_{\rm cut}$}\fi}
\newcommand{\jetset}{{\sc Jetset}}
\newcommand{\herwig}{{\sc Herwig}}
\newcommand{\jade}{{\sc Jade}}

\section{Introduction}
\label{sect-intro}
Isolated hard photon production in hadronic Z decays provides 
information about the electroweak couplings of quarks and about 
QCD~\cite{theory}. 
In \epem\ collisions at the Z resonance, initial state radiation 
is suppressed and the production of isolated photons is mainly due 
to final state radiation (FSR) from the primary quark-antiquark pair.

The aim of the analysis of hadronic events with FSR presented in
this paper is to measure the electroweak couplings to up- and down-type quarks.
The measurement is based on the photon's property of coupling to the electric charge.
By selecting hadronic Z decays with FSR 
a subsample can be extracted which is enriched in up-type quarks 
since the strength of the coupling is proportional to the charge squared.
Combining the measured rates of hadronic events with final 
state photon radiation with the total rate of hadronic Z decays
permits the simultaneous determination of the partial decay widths into
up-type and down-type quarks.
This approach is complementary to analyses of partial decay widths dealing with
either heavy or light quark flavour tagging. This analysis determines the
electroweak coupling of the up-type quarks u, c and the down-type quarks d, s, b 
for the flavour admixture present on the Z resonance.
The LEP experiments have already published QCD 
measurements
and electroweak quark coupling 
measurements
based on final state photon production~\cite{opal-qcd, EPJC2-98-39, ZPC54-92-193, aleph-qcd, delphi-qcd, l3-qcd,ZPC58-93-405, PLB301-93-136, ZPC69-95-1}.
This analysis is based on a larger data
sample and updates earlier OPAL measurements~\cite{ZPC58-93-405}.

The Z decay into a quark-antiquark pair can be described by effective
vector and axial-vector couplings, $\bar{g}_{\rm V,q}$ and 
$\bar{g}_{\rm A,q}$, where q=u or d,
representing up-type or down-type quarks. The Z decay widths
depend on the sums of the squares of these couplings. Writing 
effective couplings $c_{\mathrm q}$ as
\begin{equation}
c_{\rm q} = 4 \left( \bar{g}^2_{\rm V, q} + 
                          \bar{g}^2_{\rm A, q} \right),
\end{equation}
the total hadronic width of the Z, which is calculated to third order in
$\alpha_{\mathrm s}$ ~\cite{Gorishnii:1990vf}, is given by:

\begin{equation}
\label{eq-gamma-had}
\Gamma_{\rm hadrons} = \frac {N_{\mathrm c} m^3_{\rm Z}G_{\mathrm F}}
                         {24\pi\sqrt{2}}
    \left(1 + \frac{\alpha_{\mathrm s}}{\pi} + 1.4 \frac{\alpha_{\mathrm s}^2}{\pi^2} -12.8 \frac{\alpha_{\mathrm s}^3}{\pi^3}\right)
      (2c_{\rm u} + 3 c_{\rm d}),
\end{equation}
where $\alpha_{\mathrm s}$ is the strong coupling constant, $G_{\mathrm F}$ is the Fermi 
constant, $m_{\rm Z}$ is the mass of the Z, and $N_{\mathrm c}=3$ is the 
number of colours.  
The electromagnetic quark couplings also enter the hadronic partial width 
of the Z with a final state radiation (FSR) photon;
\begin{equation}
\label{eq-gamma-had+gamma}
  \Gamma_{{\rm hadrons}+\gamma} = \frac {N_{\mathrm c} m^3_{\rm Z}G_{\mathrm F}}
                                                         {24\pi\sqrt{2}}
    F(y_{\rm cut})\frac{\alpha}{2\pi}
      (2q_{\rm u}^2 c_{\rm u} + 3 q_{\rm d}^2 c_{\rm d}),
\end{equation}
where $\alpha$ is the electroweak coupling constant, $q_{\rm u}$ and 
$q_{\rm d}$ are the charges of up-type and down-type quarks,  
$F(y_{\rm cut})$ 
is a correction factor determined from perturbative calculations,
and $y_{\rm cut}$ is a jet resolution parameter, which determines whether a 
photon is merged into a nearby jet.
These widths can be re-written in terms of decay widths of the up- and
down-type quarks:
\begin{equation}
\label{eq-pgam-had}
  \Gamma_{\rm hadrons} = 
       2\Gamma_{\rm u} + 3\Gamma_{\rm d};~
  \Gamma_{{\rm hadrons}+\gamma} = 
       \frac{\alpha}{18\pi}\frac{F(y_{\rm cut})}{(1 + \alpha_{\mathrm s}/\pi + 1.4 \alpha_{\mathrm s}^2/\pi^2 -12.8 \alpha_{\mathrm s}^3/\pi^3)}
       (8\Gamma_{\rm u} + 3\Gamma_{\rm d}).
\end{equation}
The measurement of the partial hadronic width of the Z with an isolated 
FSR photon and the measurement of the total hadronic width of the Z depend
on different linear combinations of $c_{\rm u}$ and $c_{\rm d}$, or 
equivalently of 
$\Gamma_{\rm u}$ and $\Gamma_{\rm d}$. In order to extract the partial widths or 
the couplings $c_{\rm u,d}$, the factor $F(y_{\rm cut})$ in 
equation~\ref{eq-gamma-had+gamma} has to be calculated.
Matrix element
calculations at $\cal{O}(\alpha\alpha_{\rm s})$
are available to perform these calculations as a function of
the jet resolution parameter $y_{\rm cut}$~\cite{ZPC60-93-613}.  One important
aspect of these predictions is the regions of phase space in which they are
valid.  In particular,
infrared divergences due to soft photons and photons collinear with soft 
quarks must be avoided.  For this reason, hard isolated photons are 
studied.  A good agreement between hard isolated photon production and
matrix element predictions has already been demonstrated by 
OPAL~\cite{ZPC67-95-15}. 
In the range of intermediate values of $y_{\rm cut}$, $F(y_{\rm cut})$ can be 
determined with a theoretical uncertainty of around 1\% based on these predictions. 
The uncertainty is 
evaluated by varying $\alpha_{\rm s}$ and the phase space cut-off in the perturbative
calculation as described in more detail in section~\ref{sect-results}.
In contrast to the procedure followed in reference 
\cite{PLB301-93-136} no correction to account for the 
b-quark mass is included in the equations (\ref{eq-gamma-had}) and 
(\ref{eq-gamma-had+gamma}).
The results of the present analysis were
extracted at effective masses of the photon-jet system where the 
relative impact of the b-quark mass is 
much smaller
than in the case of reference \cite{PLB301-93-136}.

In this paper, we present a new measurement of the partial widths 
$\Gamma_{\rm u}$ and $\Gamma_{\rm d}$. The full LEP1 data sample taken
at the peak of the Z resonance was used in this analysis, amounting to 
more than 3 million hadronic Z decays.
After describing the main characteristics of the OPAL detector in 
section~\ref{sect-detector}, a description of the event selection 
follows in section~\ref{sect-selection}.  
The background and efficiency corrections applied to the selected data
are described in section~\ref{sect-bkcorrec}.
The systematic errors 
are discussed in section~\ref{sect-syst} before the results are 
presented in section~\ref{sect-results}.

\section{The OPAL Detector} 
\label{sect-detector}
The OPAL detector operated at the LEP \ee~collider at CERN.  A detailed
description of the detector can be found in reference~\cite{OPALNIM}. 
For this study, the most important components of OPAL were the central 
detector and the barrel electromagnetic calorimeter.
The central detector, measuring the momentum of charged 
particles, consisted of a system of cylindrical tracking chambers 
surrounded by a solenoidal coil which produced a uniform axial magnetic 
field of 0.435~T along the beam axis~\footnote{In the OPAL coordinate 
  system, the $x$ axis points towards the centre of the LEP ring, the 
  $y$ axis points approximately upwards and the $z$ axis points in 
the direction of 
  the electron beam.  The polar angle $\theta$ and the azimuthal angle 
  $\rm \phi$ are defined with respect to the  $z$ and $x$-axes while 
  $r$ is the distance from the $z$-axis.}.  

The electromagnetic calorimeters completely covered the azimuthal 
range for polar angles satisfying $|\cos\theta|<0.98$, providing 
excellent hermeticity. 
    The barrel electromagnetic calorimeter covered 
the polar angle range $|\cos\theta|<0.82$. It 
consisted of 9440 lead glass blocks, each 24.6 radiation length deep, 
approximately pointing towards the interaction region.  Each block subtended an 
angular region of approximately $40\times40~\rm{mrad}^{2}$. Half of the 
block width corresponded to 1.9 Moli\`ere radii. Deposits 
of energy in adjacent blocks were grouped together to form clusters
of electromagnetic energy.  The intrinsic energy resolution 
of $\rm {\sigma_E/E=0.2\%\oplus6.3\%/\sqrt{E}}$
was substantially degraded (by a factor $\simeq 2$)
by the presence of  two radiation lengths of 
material in front of the lead glass. For the    
intermediate region, $0.72<|\cos\theta|<0.82$, the amount of material 
increased to up to eight radiation lengths.
The two endcap calorimeters, each made of 
1132 lead glass blocks, 22 radiation lengths deep, covered the region of 
$0.81<|\cos\theta|<0.98$. 
In this study the measurement of inclusive photon production 
is restricted to the barrel part of the detector.    

\section{Event selection and reconstruction}
\label{sect-selection}

The first stage of the event selection was to identify hadronic Z decays.  
In this sample, isolated, high energy, neutral clusters in the
electromagnetic calorimeter were then considered as photon candidates. To
reduce the dominant background from $\pi^0$ decays, the shower shapes were
additionally required to be consistent with a shower produced by a single
photon, as opposed, for example, to overlapping showers from the two photons
from a $\pi^0$ decay. Jets were then reconstructed in events with a photon
candidate. The jet reconstruction implied an additional isolation
requirement on the photons, as a function of the jet resolution parameter.

Hadronic Z decays were selected from the data taken by
the OPAL detector between 1990 and 1995 at a centre-of-mass energy
of 91.2~GeV.  The data taken above and below the Z resonance 
during those years were not analysed.  Using the criteria described 
in reference~\cite{ZPC52-91-175}, 
$N_{\mathrm {{Z} \rightarrow Hadrons} }=$~3\,022\,897 hadronic Z 
decay candidates were selected.
The tracks and clusters were defined as described in 
reference~\cite{EPJC2-98-213}.

Isolated electromagnetic clusters in the electromagnetic calorimeter 
were identified within the selected sample.  The following criteria 
were used:
\begin{enumerate}
  \item{Only clusters with no associated tracks were considered.}
  \item{The clusters had to be located in the barrel region
        of the electromagnetic calorimeter
        defined by $|\cos\theta_{\rm cluster}|\leq 0.72$.
        This requirement reduces
        the background from initial state radiation, which
        peaks at low polar angle.
       }
  \item{The energy of each cluster was required to  be larger 
        than 7~GeV.  This reduces the background from 
        neutral hadron decays and initial state radiation.
        It also enabled the use of a shower shape fit, for 
        which the description of the Monte Carlo was not 
        satisfactory below 7~GeV.  
       }
  \item{The shape of the cluster was required to be consistent
        with a single photon.  The number $N_{\rm blocks}$
        of lead glass blocks contributing more than 0.02~GeV
        to the cluster energy was required to be less than 16.
        A cut at $W<30$~mrad was set on the energy weighted first 
        moment of the cluster, defined as:\\
        \begin{equation}
           W = \sqrt{ \frac{\sum_{i=1}^{N_{\rm blocks}} E_{i}^{\rm meas}
                              \left( (\phi_i-\bar{\phi})^2
                                    + (\theta_i - \bar{\theta})^2 \right)  }
                           {\sum_{i=1}^{N_{\rm blocks}} E_{i}^{\rm meas}} },
        \end{equation}
        where $E_{i}^{\rm meas}$, $\phi_i$ and $\theta_i$ are the 
        energy deposited in block $i$, its azimuthal and polar angles. $\bar{\phi}$ and
        $\bar{\theta}$ are the $\phi$ and $\theta$ positions of the cluster centroid.
        In addition, the shower-shape parameter $C$
        was required to be smaller than 5.
        The same definition of $C$ as in~\cite{ZPC28-85-343} was used:
\begin{equation}
\label{eq_C}
C= \frac{1}{N_{\rm blocks}}\sum_{i}\frac{(E_{i}^{\rm pred}-E_{i}^{\rm meas})^{2}}
      {(\sigma_{i}^{\rm pred})^{2}+(\sigma_{i}^{\rm meas})^{2}},
\end{equation}
where 
$E_{i}^{\rm pred}$ is the predicted energy in calorimeter block number $i$.
This was taken from the best fit of the shower profile 
parametrisation to the observed energy sharing between the calorimeter blocks
assuming that the cluster was produced by an isolated photon. The reference 
profiles varied as a function of $\cos\theta$ because of the varying amount 
of material in front of the calorimeter. 
$\sigma_{i}^{\rm pred}=0.05 \cdot E_{i}^{\rm pred}$ is the error on $E_{i}^{\rm pred}$ 
where 0.05 is 
the accuracy of the Monte Carlo integration and
$\sigma_{i}^{\rm meas}=0.002 \cdot E_{i}^{\rm meas}\oplus0.063 \sqrt{E_{i}^{\rm meas}}$ 
is the estimated error on $E_{i}^{\rm meas}$~\cite{OPALNIM}.

       }        
  \item{The cluster candidates were required to be isolated.  
        Clusters with tracks or
        other clusters of an energy sum greater than 0.5~GeV within a cone
        of half opening angle of 0.235~rad around the cluster were rejected.  
       }
\end{enumerate}

In total, 12\,626 events with a photon candidate were selected from 
the original sample of hadronic Z decays.  Each event 
in the sample was reconstructed into jets.  Two jet finding schemes 
were used and compared: \jade\ E0~\cite{jet-jade} and Durham~\cite{jet-durham}.
Both are based on the resolution parameter 
\begin{equation}
y_{ij}=\frac{M^2_{ij}}{E^2_{\mathrm vis}}
\end{equation}
where $E_{\mathrm vis}$ is the total visible energy in the detector. In the \jade\ algorithm
$M_{ij}$ is defined as the invariant mass
\begin{equation}
M_{ij}^{2~\mathrm{\jade}}=2E_{i} E_{j} (1-\cos \alpha_{ij}),
\end{equation}
while in the Durham or $k_T$ algorithm it is defined as the minimum transverse momentum 
of one particle with respect to the other,
\begin{eqnarray}
M_{ij}^{2~\mathrm{Durham}} & = & 2{\mathrm min}(E_{i}^2,E_{j}^2)(1-\cos \alpha_{ij}),
\end{eqnarray}
where $E_{i}$, $E_{j}$ are the energies of the particles and $\alpha_{ij}$ is the
angle between them.
In each case, the jet finding was performed in two steps.  In the first
step, the jets were reconstructed excluding the photon candidate 
by iteratively combining particles until for all possible particle combinations 
the calculated $y_{ij}$ was larger than a specified $y_{\rm cut}$. This was done
for 12 values of \ycut\ in both algorithms.
In the second step, $y_{\gamma i}$ was calculated for the photon
candidate and each jet~$i$ in the event.  
Each value of $y_{\gamma i}$ was required to be greater 
than the $y_{\rm cut}$ used during the first step.  
Events failing this requirement were rejected. The number of events $n_{\rm raw}$ 
passing this last selection criterion depends on the chosen $y_{\rm cut}$ value.   
The choice of values for the \ycut\ in each algorithm ensured that the photon candidate was 
sufficiently isolated from the jets and provided a scan of the phase space 
to allow a valid and more detailed comparison of the result with matrix element 
calculations. The selected sample size for different
$y_{\rm cut}$ values is shown in table~\ref{tab-n_vs_ycut} for the 
two jet finding schemes.

\section{Background and efficiency corrections}
\label{sect-bkcorrec}

The number of selected events was corrected for background 
contributions, detector response and selection efficiency. 
Two categories of background were considered. The first was the
contribution from misidentified neutral hadrons and their decay products.  
The second was from genuine photons originating from initial state
radiation. The sequence in which the data were corrected for background
and efficiency reflects this difference in background sources. First, the
background contribution from neutral hadrons was estimated by a fit to the
shower shape distribution in data, using the difference in shower shape to
distinguish genuine photons from hadron background. 
Second, a correction for detector effects was applied. 
In particular this corrected for
the photon detection efficiency and compensated for the requirements 
on the shower shape and measurement uncertainties.
It was calculated using a full simulation of the OPAL detector.
Third, events with photons produced by ISR were subtracted. Their number was estimated 
with Monte Carlo simulations at the hadron level.
Finally, the number of events remaining at the hadron level was corrected for 
hadronisation and selection efficiencies, so that it could be compared with 
results from matrix element calculations at the parton level. In this analysis,
hadron and parton level were defined as in reference \cite{dijet}.

\subsection{Neutral hadrons}
\label{sect-nhbkcorrec}
\begin{figure} [t]
  \flushleft{ \epsfig{file=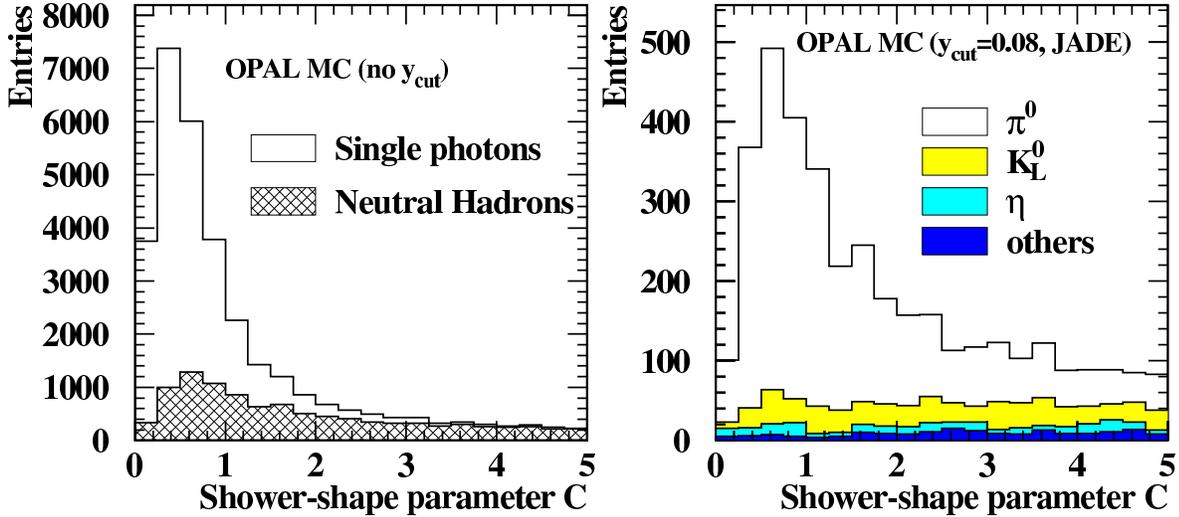,totalheight=0.29\textheight}
\caption{ \label{fig-c_dist_mc} The left plot shows distributions of the shape variable $C$ 
for single photon and neutral hadron clusters as obtained with the OPAL detector simulation 
program. The right plot shows the simulated $C$ distributions for various hadronic background 
sources in the case of the \jade\ scheme for a value of $\ycut =0.08$.}}
\end{figure}

\begin{figure} [t]
  \flushleft{ \epsfig{file=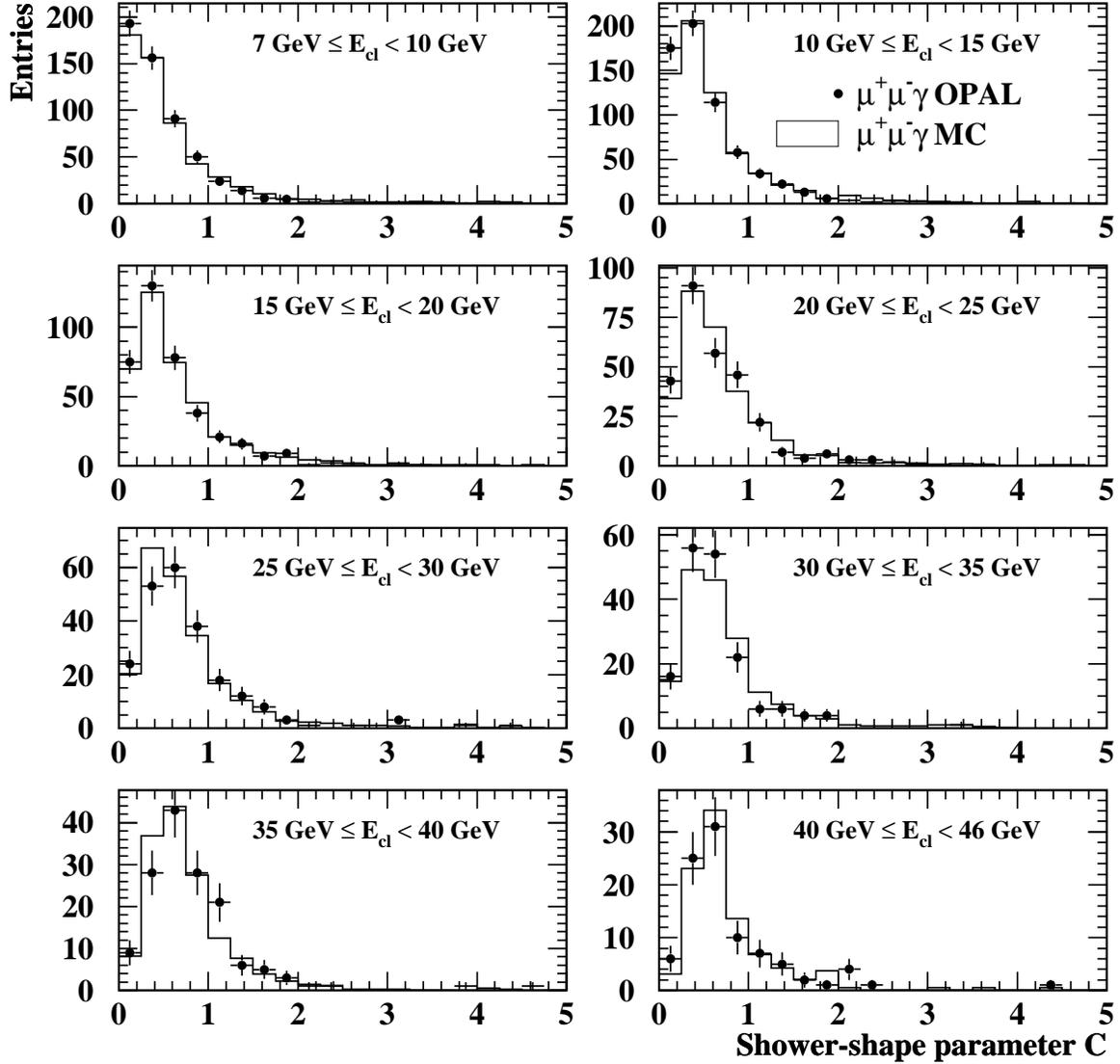,totalheight=0.725\textheight}
\caption{ \label{fig-c_dist_mu} The distributions of the shape variable $C$ for 
photon clusters in $\mu^+\mu^-\gamma$ events for data and Monte Carlo 
classified in eight bins according to their measured energy. 
The Monte Carlo distributions are 
normalised to the data statistics.
}  }
\end{figure}

\begin{figure} [htb]
  \center{ \epsfig{file=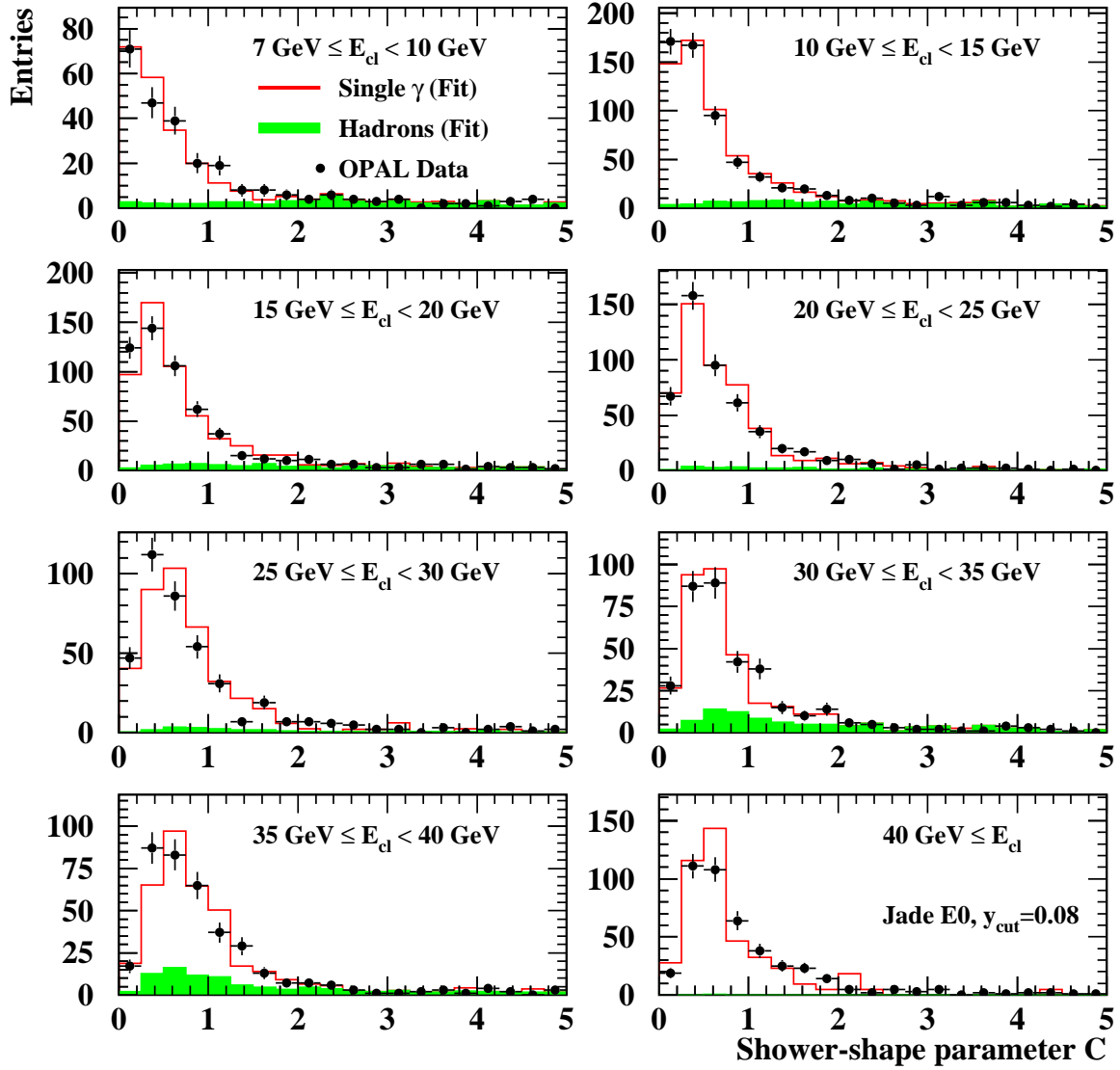,totalheight=0.725\textheight} 
\caption{ \label{fig-c_dist_fit} The data distributions of the shower-shape parameter 
$C$ in eight bins according to the measured cluster energy and the corresponding
fit result in each bin. The plots show the results in the \jade\ E0 scheme
for $\ycut =0.08$.
The filled histograms represent the distributions of the
neutral hadron background scaled to the fractions resulting from the fit.
The open area bordered by the line represents the single photon
contribution on top of the hadronic background. Note that for $E_{\rm cl}~>$~40~GeV (last plot) 
the best fit only yielded a very small contribution from hadronic background.
}  }
\end{figure}

Neutral hadrons and their decay products can produce isolated clusters 
in the electromagnetic calorimeter which pass the selection for photon 
candidates.  This background was simulated using a sample of 8.4 million 
events generated with \jetset~7.4~\cite{JETSET}, which were processed with the 
OPAL detector simulation program~\cite{GOPAL}.  The hadronic background 
comes mainly from events with a $\pi^0$ decay, and has a small 
contribution from $\eta$ decays and K$^0_L$.
There is also a contribution below 1\% of the total hadronic background 
from neutrons, $\omega$ and $\rho$.
The simulated $C$ distributions are shown in figure~\ref{fig-c_dist_mc} for
single photon clusters and for clusters produced by the hadronic background
sources. 
By fitting the data distribution to a linear combination of the expected 
distribution of the shower-shape parameter $C$ for the background and 
the measured distribution for single photons taken
from a data sample of $\mu\mu\gamma$ events, 
the fraction of single photon events was extracted~\cite{EPJC2-98-39}.
In figure \ref{fig-c_dist_mu} the $C$ distributions are plotted for isolated
neutral clusters from $\mu\mu\gamma$ events in data and Monte Carlo and
good agreement is observed.
The fit is a binned maximum likelihood fit which takes into account the effects
of data and Monte Carlo statistics~\cite{Barlow:dm}.
For each $y_{\rm cut}$ value considered, the fit was performed in eight 
cluster energy bins to extract the fraction of single photon events.  
The total fraction $f_{\gamma}(\ycut)$ of single photon events for each 
$y_{\rm cut}$ value was estimated by calculating the weighted 
average of the results for the eight 
energy bins.  The procedure was repeated for the two jet finding algorithms.  
Figure~\ref{fig-c_dist_fit} shows an example of the fit results for the 
eight cluster energy bins.  The jet finding scheme was in this case 
\jade\ E0, with $y_{\rm cut}=0.08$.
Table~\ref{tab-all_main_ycut} shows the extracted fraction of single
photon events in the selected samples for three of the $y_{\rm cut}$ values in each 
jet finding scheme, spanning the phase space studied.
The quoted $68.3\%$ confidence level error was obtained by 
studying 
the variation of $\ln{\mathcal L}$~\cite{Barlow:dm}. 
It combines the
uncertainty due to limited data statistics (2526 $\mu\mu\gamma$ events,
and between 1900 and 9000 candidates, see table~\ref{tab-n_vs_ycut})
and due to limited Monte Carlo statistics (between 548 and 1695 
clusters from neutral hadrons). 

The background estimate was 
found to be consistent with previous measurements performed by OPAL and L3.  OPAL 
measured~\cite{EPJC5-98-411} that \jetset\ underestimates 
the rate of isolated $\eta$ with an energy $E_{\eta}>4.5$~GeV by a
factor $2.07\pm 0.11$, and the rate of isolated $\pi^0$ in the energy
range $4.5<E_{\pi^0}<22.8$~GeV by a factor $1.99\pm 0.05$.  L3
measured~\cite{PLB292-92-472} a rate of isolated electromagnetic 
clusters which was $1.88\pm 0.08$ times larger than predicted by \jetset.
Applying the present method to our data sample over a range of cluster
energies $4.5 < E_{\rm clust} < 46$~GeV, we measured 
an excess of a factor $2.05\pm 0.03$.  The value is compatible
with the previous measurements of OPAL and L3.

\subsection{Detector Effects}

After subtracting the events with an isolated neutral hadron from the 
sample of candidates, the data were corrected for detector effects:
\begin{equation}
n_{\rm had}^{\gamma}(\ycut) = n_{\rm raw}(\ycut)\times
                                   f_{\gamma}(\ycut) \times
                                   c^{\rm det}(\ycut)
\end{equation}
with $n_{\rm raw}(\ycut)$ as the number of photon events before the background
correction, and
$f_{\gamma}(\ycut)$ as the fitted single photon fraction (ISR+FSR) in the selected 
sample at detector level, which has been derived as discussed in the previous 
section.
The correction factor $c^{\rm det}(\ycut)$, which takes into account detector
inefficiencies, was 
calculated for each \ycut\ and each jet finding algorithm
using the sample of 8.4 million fully simulated \jetset\ events.
The same cuts on 
the photon energy, polar angle and isolation were applied at the 
hadron level. The 
correction factors $c^{\rm det}(\ycut)$ are shown in 
table~\ref{tab-all_main_ycut}, again for three values of \ycut\ in each jet finding 
scheme. 
The efficiency losses are mainly due to photon conversions in the two 
radiation lengths of material in front of the electromagnetic calorimeter,
e.g. the beam pipe, the central tracking system and the magnet coil. The
photon conversion rate was calculated to be around 8\%.

\subsection{Initial state radiation}
Initial state radiation photons are indistinguishable from photons
from final state radiation.  Although the energy and angular cuts on 
the electromagnetic clusters reduce the amount of initial state 
radiation, which is already small when working at the Z 
resonance, initial state photons can still make a contribution of 
several percent to the selected sample.  
This contribution was evaluated using the $\cal{KK}$ Monte Carlo~\cite{hepx9912214}
without the full detector simulation; the $\cal{KK}$ generator 
has the most accurate ISR modeling for \epem collisions.
The data which are corrected for detector effects are 
directly comparable to the $\cal{KK}$ hadron level 
predictions.
The number of FSR events at the hadron level is therefore given by
\begin{equation}
  n^{\rm FSR}_{\rm had}(\ycut) = n_{\rm had}^{\gamma}(\ycut) -
                                   b^{\rm ISR}(\ycut),
\end{equation}
with $b^{\rm ISR}(\ycut)$ as the number of ISR events as calculated with the
$\cal{KK}$ Monte Carlo program and normalized to the number 
$N_{\mathrm {{Z} \rightarrow Hadrons} }$ of selected hadronic Z decays. 
The same cuts on the photon energy, polar angle and isolation were applied 
at the hadron level for both the \jetset\ and $\cal{KK}$ Monte Carlo samples.  
The numbers of isolated initial state
photons $b^{\rm ISR}(\ycut)$ to be subtracted are shown in 
table~\ref{tab-all_main_ycut} for three values of \ycut\ in each jet finding
scheme.  

\subsection{Geometrical acceptance, isolation cone, and hadronisation}
\label{sect-correction}
In order to compare the measured final state photon rate with
the matrix element predictions, the measurements must be
corrected for the remaining acceptance cuts
on the polar angle and the cone isolation,
for hadronisation, and for fragmentation,
which affect the number of jets in the event.  A correction factor $c^{\rm par}(\ycut)$ which 
takes into account the cuts mentioned above was calculated using \jetset~7.4.
No further correction was made for the energy cut $E_{\rm clust}>7$~GeV, to avoid 
introducing additional uncertainties. The matrix element predictions were 
calculated with the same requirement, $E_{\gamma}>7$~GeV.
The correction factors $c^{\rm par}(\ycut)$ for three different
\ycut\ values in two jet finding algorithms are shown in
table~\ref{tab-all_main_ycut}.  The corrected numbers of isolated final 
state photon candidates per 1000 hadron events,
\begin{equation}
   R_{\rm FSR}= n^{\rm FSR}_{\rm par}(\ycut) \times \frac{1000}{N_{\mathrm {{Z} \rightarrow Hadrons} }} \\
= n^{\rm FSR}_{\rm had}(\ycut) \times
                      c^{\rm par}(\ycut) \times \frac{1000}{N_{\mathrm {{Z} \rightarrow Hadrons} }},
\end{equation}
are shown in table~\ref{tab-n_vs_ycut}.

\section{Systematic Effects}
\label{sect-syst}
The contributions to the systematic uncertainty are summarized 
in table~\ref{tab-allsyst}. 
Relative errors are given for the value of \ycut\ in each scheme which minimizes the 
total uncertainty on the photon rate, and for the extreme \ycut\ values considered.
The statistical uncertainty quoted in table~\ref{tab-allsyst} combines
the error from limited data and Monte Carlo statistics. It is calculated
as the quadrature sum of the fit error (see section~\ref{sect-nhbkcorrec}) and
of the uncertainties of all Monte Carlo samples generated and used to
simulate various other aspects of the analysis. 
The systematic studies from which the other uncertainties were 
obtained are described below.

%
\subsection{Background}
The estimate of the hadronic background which results from the fit
of the shower-shape parameter $C$ relies on the proper simulation
of the shower for the background, as
well as on the simulation of the yield of the various hadronic background
sources.
To quantify the systematic uncertainty originating from the description
of the hadronic background, the fit procedure was repeated for 
all \ycut\ values and the two jet finding algorithms after changing
the Monte Carlo input distributions to which the data are fitted 
by varying the composition of the hadronic background. The contribution
from $\pi^0$ and $\eta$ was doubled to account for the
observed underestimation of isolated neutral pions and $\eta$-mesons in
the Monte Carlo~\cite{EPJC5-98-411}. The contribution from K$^0_L$ was
increased by 18\%, which was extrapolated from the observed differential 
cross-section in \cite{ZPC67-95-389}. 
The observed differences from the results of the default procedure 
were not statistically significant, and showed no 
systematic trend, so no systematic uncertainty was assigned.
Further systematic studies have been made by varying the
bin size and the upper boundary of the $C$ distribution.
Again, no significant change in the resulting rate was observed.

The effect of the interference between 
initial state radiation and final state radiation was checked
by comparing prompt photon samples from the $\cal{KK}$ 
Monte Carlo generated with and without ISR-FSR interference.  
The average difference of around 1 per mill in the rate of isolated 
prompt photons with an energy of larger than 7~GeV was assigned as 
systematic uncertainty on the corrected rate and is listed in 
table~\ref{tab-allsyst}.

\subsection{Detector description}
Several systematic effects coming from the description of the detector
were investigated:
\begin{itemize}
  \item{
        The track parameters were smeared by 10\% to estimate
        the effect of the reconstruction on the jet finding and the
        isolation criteria~\cite{Abbiendi:1998eh}.
       }
  \item{The effect of the energy resolution of the electromagnetic 
        calorimeter was evaluated by smearing the energy of clusters
        by 20\%. This number is chosen since the energy resolution
        of isolated photon clusters selected in \jetset\ and in a
        $\mu\mu\gamma$ reference data sample differ by up to 20\%.
        The cluster energies were also shifted by
        0.1-0.2~GeV.  This shift reflects the difference in the 
	mean values of the distributions of $E_{\rm cluster}-E_{\gamma}$
	from the simulation and of $E_{\rm cluster}-E_{\rm kin}$ in data,
	where $E_{\rm kin}$ is the kinematically reconstructed
	energy of the isolated photon in the reference
	sample.
       }
\end{itemize}
These effects would modify the correction factor $c^{\rm det}(\ycut)$.
In all cases, the systematic uncertainty on the correction factor was
found to be below the per mill level.  

An important contribution to the systematic error comes from the
simulation of photon conversions.  An error of 10\% on the photon
conversion rate~\cite{ZPC65-95-47} results in an error on the correction 
factor of about 0.8\%. 

The jet reconstruction was also tested for systematic effects.
The energy resolution was checked using a subsample of the candidate
events with two jets and
one isolated photon. 
The measured energy of each jet was compared
to the energy of the same jet calculated by imposing energy/momentum 
conservation in the final state.
The measured shift in the jet energy was used to estimate the
systematic error on the photon rate.
A variation between 0.1\% and 0.6\%
in the rate of photons was found for the \jade\ E0 jet finding scheme
as a function of $y_{\rm cut}$, and between 0.1\% and 0.3\% for the Durham scheme.

The jet angle reconstruction was tested using the same sample
of two-jet events.  The hadronic part of each event in both Monte Carlo
and data was boosted into its centre of mass using the photon momentum.
In this system the two jets should be collinear.  The angular
shift between the Monte Carlo expectation and the data was taken
to calculate a systematic uncertainty.  This uncertainty varied
between 0.1\% and 1.7\% for both jet finders. Studies of the photon 
angular resolution showed that the effect on the rate was negligible.

The combined systematic uncertainties obtained from these studies are
given in table~\ref{tab-allsyst} as the contribution to the error from the 
determination of the correction factor $c^{\rm det}(\ycut)$.

\subsection{Fragmentation and hadronisation}
Three methods were used to evaluate the uncertainty arising from fragmentation and
hadronisation.
They affect the determination of the correction factor $c^{\rm par}(\ycut)$, which compensates 
for the geometrical acceptance, i.e.\ the extrapolation of the result to the total $\cos\theta$
range, and for the isolation cone around the photon candidate. An uncertainty is not trivial to
assign since we largely rely on Monte Carlo studies.

\begin{itemize}
  \item{{\bf Parton shower model}\\
        The \jetset\ predictions were compared with those of
        \herwig\cite{HERWIG} and {\sc Ariadne}~\cite{ariadne}, which use different
        parton showering schemes.  \herwig\ also uses a different
        fragmentation scheme. 
The relative deviations are larger towards the ends of the \ycut~range
with a largest deviation of $-9.0\%$ (for \herwig\ and the Durham algorithm). 
The
deviations are smaller in the intermediate range of \ycut\ values for both 
jet finder schemes. The deviations 
stem from the different particle flow around the 
FSR photons in the three models. As a result, in the \jetset\ case more photons
are rejected at the hadron level than in the \herwig\ or {\sc Ariadne} cases 
due to the 
cone isolation requirement. This effect is smaller in the region
of intermediate \ycut\ values.
       }

  \item{{\bf First derivative prediction}\\
The photon rates measured with different \ycut\ values are correlated.
The first derivative 
of the photon rate is almost statistically independent for the different 
\ycut\ values.  Comparing the first derivative distribution for different
parton shower models provides an additional systematic cross-check
of the efficiency correction.
The first derivative is defined as
\begin{eqnarray}
D_{\rm FSR}(y_{\rm cut}) & = & \frac{1}{N_{\mathrm {{Z} \rightarrow Hadrons}}\delta}
               [N_{\rm FSR}(y_{\rm cut}-\delta/2)-N_{\rm FSR}(y_{\rm cut}+\delta/2)] \nonumber \\
                 & = & \frac{1}{N_{\mathrm {{Z} \rightarrow Hadrons}}\delta}
               [\sum_i N^{i\rightarrow {\rm reject}}-\sum_i N^{{\rm reject}\rightarrow i}],
\end{eqnarray}
where $N_{\rm FSR}$ is the number of final state photon candidates.
$N^{i\rightarrow {\rm reject}}$ is the number of 
events that have a jet multiplicity $i$ at $\ycut-\delta/2$ but that are 
rejected at $\ycut+\delta/2$, whereas $N^{{\rm reject}\rightarrow i}$ 
is the number of
events that are rejected at $\ycut-\delta/2$ but are retained at 
$\ycut+\delta/2$ with a jet multiplicity $i$. The gains and losses for
each $i$-jet event class are statistically independent, so that the
quantities,
\begin{equation}
  c^{i\rightarrow {\rm reject}}=
    \frac{N_{\rm parton}^{i\rightarrow {\rm reject}}}
         {N_{\rm hadron}^{i\rightarrow {\rm reject}}} 
     \qquad{\rm and}\qquad 
  c^{{\rm reject}\rightarrow i}=
    \frac{N_{\rm parton}^{{\rm reject}\rightarrow i}}
         {N_{\rm hadron}^{{\rm reject}\rightarrow i}}
\end{equation}
are also almost statistically independent.
The discrepancy between these quantities calculated for \herwig\ 
and \jetset\ is largest for small \ycut\ values.
Relative to the \jetset\ value, it amounts to up to --18.5\% for the
\jade\ E0 scheme and up to --24.1\% in the Durham scheme. For 
$\ycut >0.02$, the predictions agree to within 6\% for the \jade\ E0 scheme.
For the Durham scheme, they agree within 10\% for $\ycut >0.008$. 
The deviations are also due to the cone isolation criterion.
       }

  \item{{\bf Isolation cone}\\
        The analysis was repeated with different isolation cones, varying
        between 0.175~rad and 0.335~rad and allowing for up to 0.5~GeV
        of additional energy within the isolation cone.
The largest deviations are observed for a cone of $0.175$~rad half angle.
For the \jade\ E0 jet finder, they are below $5\%$ for all values of 
$\ycut$ and have a
peak value of $4.9\%$ for $\ycut=0.2$. The deviations calculated in the
Durham scheme are at the same level, with the highest value of $4.8\%$ 
at the largest $\ycut=0.1$. 
       }
\end{itemize}

Because the previous considerations are not independent, the estimates which 
give the largest positive and the largest negative contributions to the 
systematic uncertainty for each \ycut\ have been
assigned as systematic errors on the photon rate. They are listed in 
table~\ref{tab-allsyst} as the error contribution from the determination 
of $c^{\rm par}(\ycut)$ and they are typically largest for small \ycut\ values,
smaller for intermediate \ycut\ values, and increase again
for large \ycut\ values.
This indicates that the correction $c^{\rm par}(\ycut)$
and therefore the final result is more sensitive to details of the energy flow around
the photon candidate for very large and very small values of $\ycut$, i.e. near the 
phase space boundaries.

\section{Results}
\label{sect-results}

\begin{figure} [htb]
  \center{ \epsfig{file=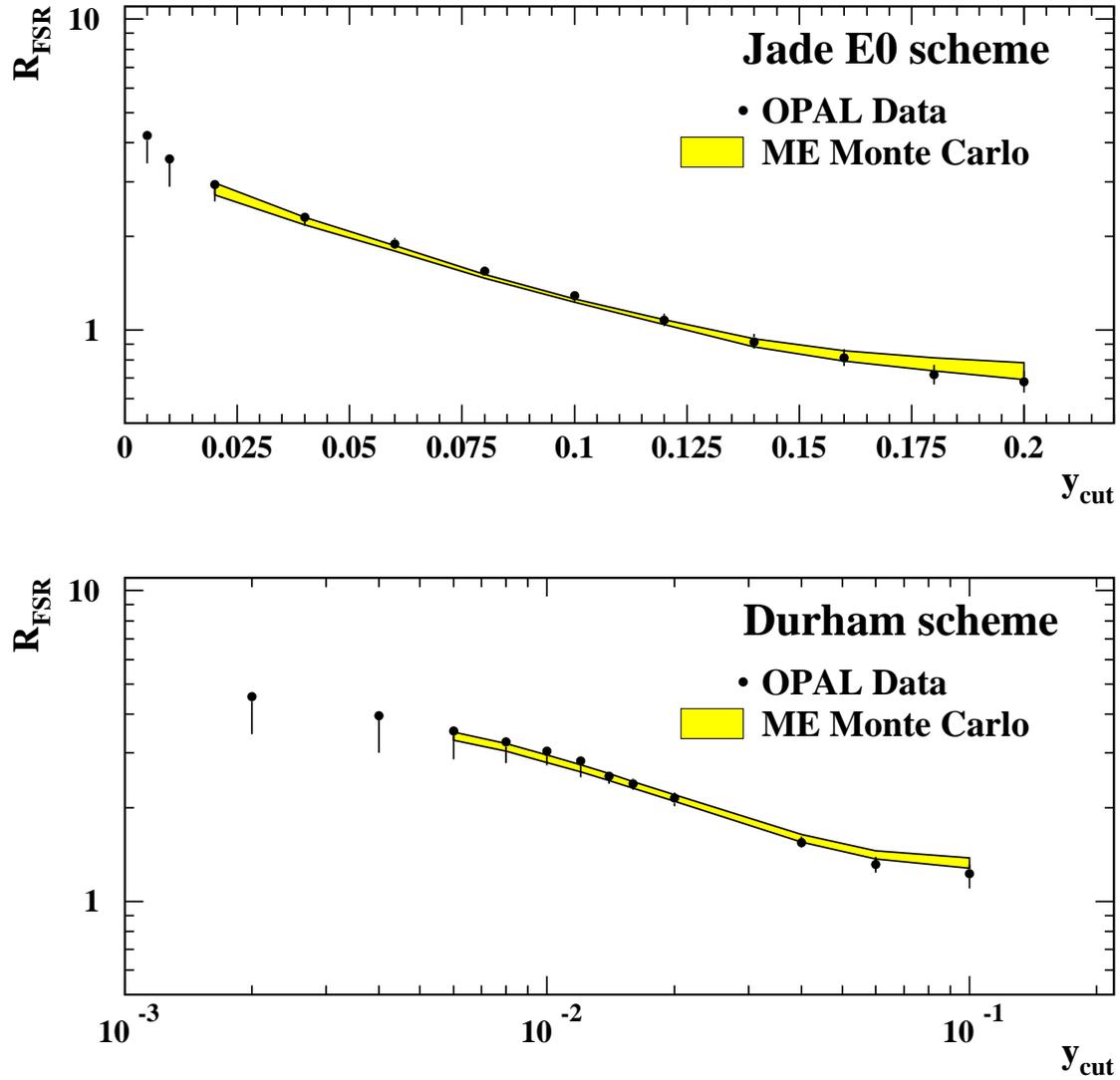, height=16.5cm} 
\caption{\label{fig-totrafig} The total number of events with FSR per 1000 
hadron events in two jet finder schemes and for twelve 
values of $y_{cut}$. The prediction of the 
${\mathcal O}(\alpha \alpha_{\mathrm s}^{(1)})$ matrix element Monte Carlo is also shown. 
The band width of the theoretical prediction is a result of a variation
of $\alpha_{\mathrm s}^{(1)}$ and of the phase space cut-off in the 
perturbative calculations.}  }
\end{figure}

Figure~\ref{fig-totrafig} shows the measured rate of FSR photons
normalised to 1000 hadronic Z decays as a function of
the \ycut\ value for the two jet finding algorithms.  
The results are compared to bands of matrix element
Monte Carlo predictions for $0.15 < \alpha^{(1)}_{\mathrm s} < 0.22$ and
$0.0005 < y_{\gamma} < 0.001$. 
$\alpha^{(1)}_{\mathrm s}$ is the first order $\alpha_{\mathrm s}$ value and
$y_{\gamma}$ is the cut-off in the
phase space of the photon emission which is used in the matrix element 
calculations to remove the singularity of the 
$y_{q\gamma}$,$y_{\bar{q}\gamma}$ poles~\cite{ZPC58-93-405,ZPC60-93-613,ZPC67-95-15,ZPC54-92-193}.  
The lowest two values of \ycut\ in both jet finder schemes
are not considered because of the large contribution from events
with more than three jets, which are not described by the
matrix element calculations.
For both jet finding algorithms, the measurement errors in the 
intermediate \ycut\ range are 
smaller than in the regions of small and large \ycut\ values.
At small and large \ycut\ values, the result is particularly sensitive 
to the isolation cone requirement, causing a large efficiency loss and a
large systematic error.

The decay widths $\Gamma_{\rm u}$ and $\Gamma_{\rm d}$ 
can be calculated
from the measured FSR photon rate.  
The value $\ycut=0.08$ for the \jade\ E0 jet finding 
algorithm and $\ycut=0.016$ for Durham were chosen for the final results
since they are the values for which the error on the FSR photon rate is
the smallest.
Taking a value of $\alpha_{\mathrm s}^{(1)} = 0.18\pm0.03$~\cite{Acton:1992fa}, correction
factors $F(y_{\rm cut})$ of 
$9.60^{+0.16}_{-0.13}$ ($\ycut=0.016$, Durham scheme) and $6.11^{+0.01}_{-0.06}$ ($\ycut=0.08$, \jade\ E0 scheme)
were calculated to be used as input to equation~\ref{eq-gamma-had+gamma}.
With values of 
$\alpha_s=0.1172\pm 0.002$ and $\Gamma_{\rm hadrons}=1744.4\pm 2.0$~MeV~\cite{pdg} 
entering equation~\ref{eq-gamma-had},
we obtain

\begin{equation}
  8\Gamma_{\rm u} + 3\Gamma_{\rm d} = 
            3.55\pm0.08^{+0.07+0.03}_{-0.07-0.01}\;\;{\rm GeV}
\end{equation}
for the \jade\ E0 jet finding algorithm ($\ycut=0.08$) and
\begin{equation}
  8\Gamma_{\rm u} + 3\Gamma_{\rm d} = 
            3.50\pm0.09^{+0.10+0.05}_{-0.14-0.06}\;\;{\rm GeV}
\end{equation}
for Durham ($\ycut=0.016$). The first error is due to statistics, the second 
error is due to systematics and the third error comes from the uncertainty in
$\alpha_{\mathrm s}^{(1)}$ and the theoretical cut off $y_{\gamma}$.
The results derived with the \jade\ and Durham schemes are consistent 
and the result with the smaller
uncertainty (\jade\ scheme) is considered the main result.

\begin{figure} [htb]
  \center{ \epsfig{file=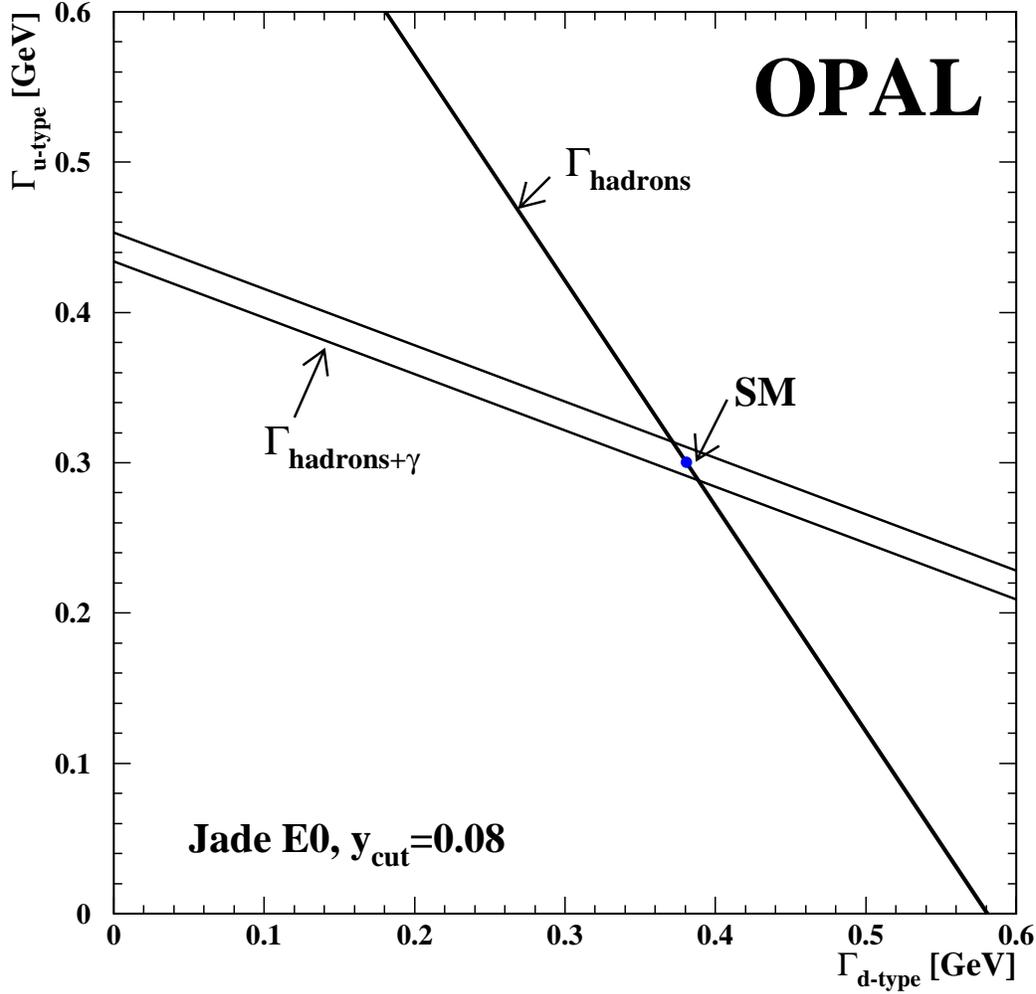, height=15cm} 
\caption{\label{fig-correlation} Correlation between the Z decay widths to 
up-type and down-type quarks. The band width 
of $\Gamma_{\rm hadrons +\gamma}= 8\Gamma_{\rm u} + 3\Gamma_{\rm d}$ reflects
one standard deviation as obtained from this measurement.
The narrow band shows the correlation within one standard deviation as obtained
from the world average of $\Gamma_{\rm hadrons} = 2\Gamma_{\rm u} + 3\Gamma_{\rm d}$.
The point on the narrow band represents the Standard Model expectation.}
}
\end{figure}

Figure~\ref{fig-correlation} shows the correlation between the Z decay
widths in up-type and down-type quarks obtained from this measurement.
Also shown are the Standard Model prediction and the correlation obtained 
from the measurement of 
$\Gamma_{\rm hadrons} = 2\Gamma_{\rm u} + 3\Gamma_{\rm d}$.  
Combining
the information on $\Gamma_{{\rm hadrons}+\gamma}$ and $\Gamma_{\rm hadrons}$, the
decay widths for up-type and down-type quarks can be extracted for
$y_{cut}=0.08$ in the \jade\ scheme:
\begin{equation}
  \Gamma_{\rm u}=300\pm 13^{+12+5}_{-12-1} \quad{\rm MeV}
  \qquad {\rm and} \qquad 
  \Gamma_{\rm d}=381\pm  9^{+8+1}_{-8-4} \quad{\rm MeV}\nonumber.
\end{equation}
Note that in this measurement $\Gamma_{\rm u}$ and 
$\Gamma_{\rm d}$ are 100\% anti-correlated. 
The Standard Model predictions~\cite{pdg} for these widths are:
\begin{equation}
  \Gamma_{\rm u}= 300.3\pm 0.14 \quad{\rm MeV}
  \qquad {\rm and} \qquad 
  \Gamma_{\rm d}= 380.7\pm 0.15 \quad{\rm MeV}\nonumber
\end{equation}
in good agreement with the measurement. 
The result is also in good agreement with
previous measurements from 
DELPHI~\cite{ZPC69-95-1}, L3~\cite{PLB301-93-136}, and OPAL~\cite{ZPC58-93-405,Ackerstaff:1997rc} 
and is significantly more precise.
The analysis in~\cite{Ackerstaff:1997rc} used high-momentum stable particles as
a tag, while the other three were based on the determination of the FSR rate in 
hadronic Z decays. 

\section{Conclusion}

Using the entire OPAL LEP1 on-peak Z hadronic decay sample 
a measurement of
the partial decay widths of the Z into up- and down-type quarks 
was performed.
The measurement was based on 12\,626 observed candidate 
hadronic Z decays with final state photon emission.
The partial width $\Gamma_{{\rm hadrons}+\gamma}$ for such decays was calculated 
as a function of the partial widths of the Z into up- and down-type 
quarks with the aid of a
matrix element Monte Carlo program of $\cal{O}(\alpha\alpha_{\rm s})$.
The calculated value for $\Gamma_{{\rm hadrons}+\gamma}$ was combined 
with the PDG average of the total hadronic 
Z width $\Gamma_{\rm hadrons}$ to determine 
$\Gamma_{\rm u}=300^{+19}_{-18}$~MeV
and $\Gamma_{\rm d}=381^{+12}_{-12}$~MeV. 
The result is consistent with the Standard Model expectation and
with previous 
measurements~\cite{ZPC58-93-405,PLB301-93-136,ZPC69-95-1,Ackerstaff:1997rc}.
Its precision is significantly improved compared to previous
measurements using a similar method of selecting hadronic Z decays with
highly isolated and energetic photons due to increased data and Monte Carlo
statistics. 
With respect to the previous OPAL measurement~\cite{ZPC58-93-405} a different approach was chosen
to estimate the background rate of neutral hadrons being misidentified as genuine 
photons. The previous measurement invokes isospin symmetry to estimate the 
background rate of neutral hadrons and measures the rate of isolated charged pions 
to extract the rate of isolated neutral pions.
The present analysis  
utilises the different shapes of the distribution of the shower-shape parameter $C$ for
neutral hadrons and genuine photons.
The larger data and Monte Carlo statistics allowed us to choose this method which 
exhibits a smaller systematic uncertainty.

\section*{Acknowledgements}
\par
We particularly wish to thank the SL Division for the efficient operation
of the LEP accelerator at all energies
 and for their close cooperation with
our experimental group.  In addition to the support staff at our own
institutions we are pleased to acknowledge the  \\
Department of Energy, USA, \\
National Science Foundation, USA, \\
Particle Physics and Astronomy Research Council, UK, \\
Natural Sciences and Engineering Research Council, Canada, \\
Israel Science Foundation, administered by the Israel
Academy of Science and Humanities, \\
Benoziyo Center for High Energy Physics,\\
Japanese Ministry of Education, Culture, Sports, Science and
Technology (MEXT) and a grant under the MEXT International
Science Research Program,\\
Japanese Society for the Promotion of Science (JSPS),\\
German Israeli Bi-national Science Foundation (GIF), \\
Bundesministerium f\"ur Bildung und Forschung, Germany, \\
National Research Council of Canada, \\
Hungarian Foundation for Scientific Research, OTKA T-038240, 
and T-042864,\\
The NWO/NATO Fund for Scientific Research, the Netherlands.\\

%

\newpage

\begin{table}
  \begin{center}
  \begin{tabular}{|c|c|c|c|c|c|}
\hline
\multicolumn{3}{|c|}{\jade\ E0} & \multicolumn{3}{|c|}{Durham ($k_t$)} \\ \hline  
$y_{cut}$ & $n_{\rm raw}$ & $R_{\rm FSR}$ & $y_{cut}$ & $n_{\rm raw}$ & $R_{\rm FSR}$      \\ \hline\hline 
  0.005 & 8503 & $ 4.22 \pm 0.06 $ &0.002 & 8936 & $ 4.55 \pm 0.07 $\\ \hline
  0.01  & 7504 & $ 3.55 \pm 0.06 $ &0.004 & 8017 & $ 3.96 \pm 0.06 $\\ \hline
  0.02  & 6351 & $ 2.93 \pm 0.05 $ &0.006 & 7331 & $ 3.53 \pm 0.06 $\\ \hline
  0.04  & 5039 & $ 2.31 \pm 0.05 $ &0.008 & 6876 & $ 3.25 \pm 0.06 $\\ \hline
  0.06  & 4140 & $ 1.89 \pm 0.04 $ &0.01  & 6366 & $ 3.04 \pm 0.05 $\\ \hline
  0.08  & 3496 & $ 1.54 \pm 0.03 $ &0.012 & 5996 & $ 2.83 \pm 0.05 $\\ \hline
  0.1   & 2904 & $ 1.29 \pm 0.03 $ &0.014 & 5613 & $ 2.53 \pm 0.08 $\\ \hline
  0.12  & 2513 & $ 1.07 \pm 0.04 $ &0.016 & 5331 & $ 2.39 \pm 0.06 $\\ \hline
  0.14  & 2261 & $ 0.91 \pm 0.04 $ &0.02  & 4890 & $ 2.15 \pm 0.06 $\\ \hline
  0.16  & 2094 & $ 0.81 \pm 0.05 $ &0.04  & 3743 & $ 1.54 \pm 0.05 $\\ \hline
  0.18  & 1991 & $ 0.72 \pm 0.05 $ &0.06  & 3288 & $ 1.32 \pm 0.06 $\\ \hline
  0.2   & 1933 & $ 0.68 \pm 0.05 $ &0.1   & 2913 & $ 1.23 \pm 0.06 $\\ \hline
  \end{tabular}
  \caption{\label{tab-n_vs_ycut} Total number of candidate events $n_{\rm raw}$ 
and corrected number of selected candidate events in 1000 
hadron events $R_{\rm FSR}$ in the two jet finder schemes for twelve different values of 
$y_{cut}$. The error represents the combined uncertainty from data and Monte Carlo
statistics.
}
  \end{center}
\end{table}

\begin{table}
  \begin{center}
  \begin{tabular}{|c|c|c|c|}
\hline
                         & \multicolumn{3}{|c|}{\jade\ E0 scheme}                    \\ \hline
$y_{cut}$                & 0.005           & 0.08              & 0.20                \\ \hline
$f_{\gamma}(\ycut)$      & $0.851 \pm 0.011$ & $0.865 \pm 0.016$ & $0.670 \pm 0.042$ \\ \hline
$c^{\rm det}(y_{cut})$   & $1.142\pm 0.005$  & $1.152\pm 0.008$  & $1.073\pm 0.011$  \\ \hline
$b^{\rm ISR}(\ycut)$     & $354 \pm 8$       & $183 \pm 6$       & $ 83 \pm 4$       \\ \hline
$c^{\rm par}(\ycut)$     & $1.612 \pm 0.008$ & $1.413 \pm 0.010$ & $1.573 \pm 0.018$ \\ \hline \hline
                       & \multicolumn{3}{|c|}{Durham ($k_T$) scheme}               \\ \hline
$y_{cut}$              & 0.002             & 0.016             & 0.1               \\ \hline
$f_{\gamma}(\ycut)$    & $0.839 \pm 0.012$ & $0.848 \pm 0.019$ & $0.719 \pm 0.035$ \\ \hline
$c^{\rm det}(y_{cut})$ & $1.146\pm 0.004$  & $1.137\pm 0.006$  & $1.062\pm 0.008$  \\ \hline
$b^{\rm ISR}(\ycut)$   & $380 \pm 8$       & $222 \pm 6$       & $109 \pm 5$       \\ \hline
$c^{\rm par}(\ycut)$   & $1.673 \pm 0.008$ & $1.471 \pm 0.008$ & $1.754 \pm 0.017$  \\ \hline

  \end{tabular}
  \caption{\label{tab-all_main_ycut} The single-photon fraction $f_{\gamma}(\ycut)$, 
the correction factor $c^{\rm det}(y_{\rm cut})$ which takes into account detector 
inefficiency, the expected number $b^{\rm ISR}(\ycut)$ of background 
events due to initial state radiation in the sample of selected candidates, and
the efficiency correction $c^{\rm par}(\ycut)$. For each jet finding scheme, values are 
shown for the \ycut\ which gives the minimum total error, and for the two extreme \ycut\ values.
The errors reflect the available Monte Carlo statistics and in the case of the 
single-photon fraction the combined data and Monte Carlo statistics. 
}
  \end{center}
\end{table}

\begin{table}
  \begin{center}
  \begin{tabular}{|c|c|c|c|c|c|c|}
\hline
\multicolumn{7}{|c|}{Relative error in \% for} \\ \hline\hline
                         & \multicolumn{3}{|c|}{\jade\ E0 scheme}      & \multicolumn{3}{|c|}{Durham ($k_T$) scheme}\\ \hline
$y_{cut}$                & 0.005    & 0.08     & 0.20     & 0.002    & 0.016    & 0.1      \\ \hline
Statistics               & $\pm$1.5 & $\pm$2.2 & $\pm$6.8 & $\pm$1.6 & $\pm$2.4 & $\pm$5.3 \\ 
(candidates, $\mu\mu\gamma$, MC)&   &          &          &          &          &          \\ \hline
Systematics              &          &          &          &          &          &          \\ \hline
ISR background           & $\pm$0.1 & $\pm$0.1 & $\pm$0.1 & $\pm$0.1 & $\pm$0.1 & $\pm$0.1 \\ \hline
$c^{\rm det}(\ycut)$     & $\pm$1.8 & $\pm$1.2 & $\pm$0.8 & $\pm$1.8 & $\pm$1.5 & $\pm$0.8 \\ \hline
$c^{\rm par}(\ycut)$     & +1.5     & +1.6     & +4.9     & +1.8     &  +2.5    & +4.8     \\ 
                         & --18.5   & --1.6    & --3.7    & --24.1   &  --3.6   & --9.0    \\ \hline
Sum                      & +2.7     & +3.0     & +8.4     & +3.0     &  +3.8    & +7.2     \\ 
                         & --18.6   & --3.0    & --7.8    & --24.2   &  --4.6   & --10     \\ \hline
  \end{tabular}
   \caption{\label{tab-allsyst} Summary of error contributions for the cases with the smallest
total uncertainties in each of the two jet finder schemes. The error contributions 
for the two extreme values are also shown. The estimation of the various contributions is
discussed in section \ref{sect-syst}.
}
  \end{center}
\end{table}


\end{document}